\documentclass[preprint,showpacs,preprintnumbers,amsmath,amssymb]{revtex4-1}
\usepackage{mathrsfs}
\usepackage{graphicx}
\usepackage{dcolumn}% Align table columns on decimal point
\usepackage{bm}% bold math
\begin{document}

%\begin{linenumbers}

\title{Quantum-mechanical description of entanglement between photon's polarization and momentum}

\author{Chun-Fang Li\footnote{Email address: cfli@shu.edu.cn}}

\affiliation{Shanghai Key Lab of Modern Optical System, Engineering Research Center of Optical Instrument and System, Ministry of Education, School of Optical-Electrical and Computer Engineering, University of Shanghai for Science and Technology, Shanghai 200093, China}

\affiliation{Department of Physics, Shanghai University, 99 Shangda Road, 200444
Shanghai, China}

\date{\today}% It is always \today, today,
             %  but any date may be explicitly specified

\begin{abstract}

It has been accepted that the polarization of the photon in vector beams is entangled with its momentum. Here a quantum description is advanced for the polarization that shows entanglement with the momentum.
This is done by showing that the Jones vector at each value of the momentum plays the role of the polarization wavefunction in the sense that the Pauli matrices represent the Cartesian components of the polarization in the local reference system with respect to which the Jones vector is defined.
The unit vector that the constraint of transversality condition requires to specify the local reference system turns out to be a gauge degree of freedom that determines the entanglement of the polarization with the momentum and has observable effects.

\end{abstract}

%\pacs{42.50.Tx, 03.65.Ca, 42.90.+m}% PACS, the Physics and Astronomy
                                   % Classification Scheme.
%\keywords{}                       % Use showkeys class option if keyword
                                   % display desired
\maketitle

%----------------------------------------------------------------

\textit{Introduction}
The Stokes parameters in classical optics describe the polarization of a plane wave \cite{Jauch, Born}. They are usually characterized as the Cartesian components of the so-called Stokes vector on the Poincar\'{e} sphere \cite{Damask}. Such parameters have been generalized to operators in quantum optics \cite{Jauch, Kara, Luis, Marq, Sode} to describe the polarization of the photon.
But on the other hand, it has been gradually recognized that the polarization of the photon in vector beams \cite{Hall, Stal, Beck, Card} is entangled with its momentum \cite{Gadw}, its position \cite{Kari}, and its orbital angular momentum \cite{Naga, Vall, D'Am}. For example, the cylindrical-vector beams \cite{Pohl, Jord-H, Youn-B, Zhan} show cylindrically symmetric polarization.
The purpose of this Letter is to put forward an operator for the polarization that carries the hybrid entanglement with the momentum.
The possibility for us to do so lies in the observation that the Stokes parameters in classical optics have a definite relation with the local reference system with respect to which the Jones vector is defined.
To our surprise, the degree of freedom to specify the local reference system determines how the polarization is entangled with the momentum and has observable effects.
Because we are concerned only with the polarization of single photons, the discussion will be made in the framework of first quantization \cite{Akhiezer}.

In quantum mechanics, the intrinsic degree of freedom of particles is described by
a multiple-component wavefunction. For instance, the spin of the electron is described by a two-component wavefunction.
It is well known that the electric vector $\mathcal{E}(\mathbf{x},t)$ of a radiation field in position space satisfies the transversality condition,
\begin{equation}\label{TC-E}
    \nabla \cdot \mathcal{E}=0.
\end{equation}
Since the oscillation direction of the electric vector reflects the polarization state of the photon, this condition implies the entanglement of the polarization with the position. But unfortunately, the electric vector is not the wavefunction of the photon.
As a matter of fact, due to the nonlocality of the photon \cite{Jauch-P, Amrein, Pauli, Rose-S}, the position-space wavefunction cannot be introduced \cite{Akhiezer, Biru} in the usual sense \cite{Sakurai}.
However, the momentum-space wavefunction is well defined. In a manifestly non-relativistic formalism \cite{Cohen}, it is a vector function $\mathbf{f}(\mathbf{k}, t)$ satisfying the Schr\"{o}dinger equation
\begin{equation}\label{SE-V}
    i \frac{\partial \mathbf{f}}{\partial t}= \omega \mathbf{f},
\end{equation}
where $\mathbf k$ is the wavevector, $\omega =ck$ is the angular frequency playing the role of the Hamiltonian, and  $k=|\mathbf{k}|$.
So the polarization, the intrinsic degree of freedom, of the photon is described quantum-mechanically by the vector wavefunction (VWF). Here there is a peculiar feature that does not usually occur in quantum processes. This is that the three components of the VWF are not independent.
They are constrained by the transversality condition,
\begin{equation}\label{TC}
    \mathbf{f} \cdot \mathbf{w}=0,
\end{equation}
where $\mathbf{w}=\mathbf{k}/k$ is the unit wavevector.
It should be pointed out that the Schr\"{o}dinger equation (\ref{SE-V}) together with the transversality condition (\ref{TC}) is equivalent to the relativistic Maxwell's equations \cite{Akhiezer, Cohen}.
In particular, when written as
$\mathcal{E}=\frac{1}{\sqrt 2}(\mathbf{E}+\mathbf{E}^\ast)$,
the electric vector of a radiation field can be expressed in terms of the VWF as
\begin{equation}\label{E-vec}
    \mathbf{E} (\mathbf{x},t)
   =\frac{1}{(2 \pi)^{3/2}} \int \Big( \frac{\hbar \omega}{\varepsilon_0} \Big)^{1/2}
    \mathbf{f} (\mathbf{k}, t) e^{i \mathbf{k} \cdot \mathbf{x}} d^3 k.
\end{equation}
In view of this, Eq. (\ref{TC}), which now implies the entanglement of the polarization with the momentum, should be regarded as a quantum-mechanical constraint that is imposed by the relativistic nature of the photon.
What is actually done in this Letter is to investigate the role of the constraint (\ref{TC}) in the quantum-mechanical description of the polarization.

\textit{From Stokes parameters to polarization operator}
Constrained by the transversality condition (\ref{TC}), the VWF has only two independent components. To see how the constraint affects the quantum-mechanical description of the polarization, it is beneficial to convert the VWF into a wavefunction that consists of two independent components.
Fortunately, such a conversion has been prepared in classical optics in the context of a plane wave, giving rise to the Jones vector \cite{Damask}. Let us first introduce the polarization operator by reexamining the property of the Stokes parameters that are completely determined by the Jones vector and appreciate the physical significance of the gauge degree of freedom that comes from the constraint (\ref{TC}).

Suppose a monochromatic plane wave the momentum of which is along the positive $z$-axis. Its electric vector can be written as
$\mathbf{E}_0 (z,t)=E \mathbf{a}_0 \exp [i(k_0 z-\omega_0 t)]$,
where $E$ denotes the amplitude and $\mathbf{a}_0$ is the unit electric vector.
Due to the transversality condition (\ref{TC-E}), the momentum defines the transverse plane within which the electric vector $\mathbf{a}_0$ is located.
As a result, $\mathbf{a}_0$ can be expanded in terms of two orthogonal base vectors that are also located within the plane.
Consider an arbitrary Cartesian reference system $xy$ in this plane and denote its axes by unit vectors $\mathbf{e}_x$ and $\mathbf{e}_y$ that form with the unit vector $\mathbf{e}_z$ of the $z$-axis a right-handed reference system $xyz$.
Choosing $\mathbf{e}_x$ and $\mathbf{e}_y$ as the base vectors as usual, $\mathbf{a}_0$ can be expanded as
$
\mathbf{a}_0=a_x \mathbf{e}_x +a_y \mathbf{e}_y.
$
The expansion coefficients make up the Jones vector
$
\tilde{a}_0=\bigg(\begin{array}{c}
                    a_x \\
                    a_y
                  \end{array}
            \bigg)
$.
For convenience, a $3 \times 2$ matrix
$
\varpi_0=(\begin{array}{cc}
            \mathbf{e}_x & \mathbf{e}_y
          \end{array})
$
is introduced to rewrite the expansion as
\begin{equation}\label{a}
    \mathbf{a}_0=\varpi_0 \tilde{a}_0,
\end{equation}
where vectors of three Cartesian components such as $\mathbf{e}_x$ and $\mathbf{e}_y$ are expressed as column matrices.
Since
$\varpi_0^\dag \varpi_0=I_2$,
where the superscript $\dag$ stands for the conjugate transpose and $I_2$ is the $2 \times 2$ unit matrix, it is seen from Eq. (\ref{a}) that the Jones vector can be expressed in terms of the electric vector as
\begin{equation}\label{JV}
    \tilde{a}_0=\varpi_0^\dag \mathbf{a}_0.
\end{equation}
With the Jones vector, the Stokes parameters of the plane wave are given by \cite{Damask}
\begin{equation}\label{SP}
    s_{0i} =\tilde{a}_0^\dag \hat{\sigma}_i \tilde{a}_0,
\end{equation}
where $\hat{\sigma}_i$'s are the Pauli matrices,
\begin{equation}\label{PM}
    \hat{\sigma}_1=\bigg(\begin{array}{cc}
                           1 &  0 \\
                           0 & -1
                         \end{array}
                   \bigg), \quad
    \hat{\sigma}_2=\bigg(\begin{array}{cc}
                           0 & 1 \\
                           1 & 0
                         \end{array}
                   \bigg), \quad
    \hat{\sigma}_3=\bigg(\begin{array}{cc}
                           0 & -i \\
                           i &  0
                         \end{array}
                   \bigg).
\end{equation}
The Stokes parameters (\ref{SP}) are completely determined by the Jones vector $\tilde{a}_0$. Nevertheless, they are not completely determined by the electric vector $\mathbf{a}_0$.
This is because the Jones vector is defined with respect to the momentum-dependent reference system $xyz$. But the momentum cannot define the transverse reference system $xy$. It defines only the transverse plane.
There is a degree of freedom to choose the transverse reference system that is different from one another by a rotation about the momentum. It is thus essential to analyze how the Stokes parameters are related to the momentum-dependent reference system $xyz$.

To this end, consider a new momentum-dependent reference system $x' y' z$ that is different from the old one $xyz$ by a rotation of an angle $\phi$ about the momentum. The unit vectors of its transverse axes are given by
\begin{eqnarray*}
% \nonumber to remove numbering (before each equation)
  \mathbf{e}_{x'} &=&  \mathbf{e}_{x} \cos \phi +\mathbf{e}_{y} \sin \phi , \\
  \mathbf{e}_{y'} &=& -\mathbf{e}_{x} \sin \phi +\mathbf{e}_{y} \cos \phi .
\end{eqnarray*}
Letting
$
\varpi'_0=(\begin{array}{cc}
             \mathbf{e}_{x'} & \mathbf{e}_{y'}
           \end{array})
$,
these two equations can be rewritten compactly as
\begin{equation}\label{varpi'}
    \varpi'_0 =\exp[-i (\hat{\mathbf \Sigma} \cdot \mathbf{e}_z) \phi ] \varpi_0,
\end{equation}
or as
\begin{equation}\label{varpi'-s3}
    \varpi'_0 =\varpi_0 \exp (-i \hat{\sigma}_3 \phi),
\end{equation}
where
$(\hat{\Sigma}_k)_{ij} =-i \epsilon_{ijk}$
with $\epsilon_{ijk}$ the Levi-Civit\'{a} pseudotensor.
In the new reference system, the Jones vector of the plane wave becomes
\begin{equation}\label{JV'}
    \tilde{a}'_0 =\varpi'^\dag_0 \mathbf{a}_0 =\exp (i \hat{\sigma}_3 \phi) \tilde{a}_0
\end{equation}
by virtue of Eqs. (\ref{varpi'-s3}) and (\ref{JV}), which in turn expresses the electric vector as
\begin{equation}\label{a+}
    \mathbf{a}_0=\varpi'_0 \tilde{a}'_0.
\end{equation}
Accordingly, the Stokes parameters become
\begin{subequations}\label{SP'}
\begin{align}
  s'_{01} &=\tilde{a}'^\dag_0 \hat{\sigma}_1 \tilde{a}'_0
           =s_{01} \cos 2\phi +s_{02} \sin 2\phi, \\
  s'_{02} &=\tilde{a}'^\dag_0 \hat{\sigma}_2 \tilde{a}'_0
           =-s_{01} \sin 2\phi +s_{02} \cos 2\phi, \\
  s'_{03} &=\tilde{a}'^\dag_0 \hat{\sigma}_3 \tilde{a}'_0=s_{03} \label{SP3}.
\end{align}
\end{subequations}
Apparently, the first two Stokes parameters depend on the choice of the momentum-dependent reference system. Only the third one does not.
It is well known that the Stokes parameters can be characterized as the Cartesian components of the Stokes vector on the Poincar\'{e} sphere.
Now that they are completely determined by the Jones vector, it is only possible to stipulate that they are the Cartesian components of a vector in the momentum-dependent reference system with respect to which the Jones vector is defined if they are to be related to this reference system. In particular, Eqs. (\ref{SP'}) suggest that the Stokes parameters (\ref{SP}) constitute the following vector in the old reference system,
\begin{equation}\label{s}
    \mathbf{s}_0=s_{01} \mathbf{e}_x +s_{02} \mathbf{e}_y +s_{03} \mathbf{e}_z
                \equiv \tilde{a}^\dag_0 \hat{\boldsymbol \sigma} \tilde{a}_0,
\end{equation}
where
\begin{equation}\label{PO}
    \hat{\boldsymbol \sigma}=\hat{\sigma}_1 \mathbf{e}_x+\hat{\sigma}_2 \mathbf{e}_y
                            +\hat{\sigma}_3 \mathbf{e}_z.
\end{equation}
So introduced vector is known in quantum mechanics \cite{Merz} as the polarization vector.
By this it is meant that analogous to the spin of the electron, the polarization of the photon can also be represented by the Pauli matrices.
But it is noted that the polarization wavefunction (PWF) here is the Jones vector $\tilde{a}_0$ rather than the electric vector $\mathbf{a}_0$. This is in consistency with the fact that the Pauli matrices represent the Cartesian components of the polarization in the momentum-dependent reference system $xyz$ with respect to which the PWF is defined.
Furthermore, because the Pauli matrices satisfy the canonical commutation relation of the angular momentum,
\begin{equation}\label{CCR}
    [\hat{\sigma}_i, \hat{\sigma}_j]=2i \epsilon_{ijk} \hat{\sigma}_k,
\end{equation}
except for a factor two, only in the momentum-dependent reference system can the polarization (\ref{s}) transform as a vector \cite{Sakurai} under the rotation that is generated by the Pauli matrices.
From now on we will only refer to this vector as the polarization.

In classical electromagnetism one can choose different gauge potentials to express the same electromagnetic wave. Here one encounters a similar situation. This is that there is a degree of freedom to choose the PWF to express the same electric vector as Eqs. (\ref{a}) and (\ref{a+}) illustrate. It appears that the PWF is a gauge representation for the polarization of the photon.
The gauge here is the momentum-dependent reference system. Once the gauge is specified, the PWF has a one-to-one correspondence with the electric vector via Eq. (\ref{JV}). Eq. (\ref{JV'}) is the gauge transformation on the PWF.
In this sense, Eq. (\ref{s}) is the polarization of the photon in the old gauge.
The polarization of the same photon in the new gauge is given by
\begin{equation*}
    \mathbf{s}'_0=\tilde{a}'^\dag_0 \hat{\boldsymbol \sigma}' \tilde{a}'_0
                 =s'_{01} \mathbf{e}_{x'} +s'_{02} \mathbf{e}_{y'} +s'_{03} \mathbf{e}_z,
\end{equation*}
where the corresponding polarization operator is
\begin{equation}\label{PO'}
    \hat{\boldsymbol\sigma}'=\hat{\sigma}_1 \mathbf{e}_{x'}+\hat{\sigma}_2 \mathbf{e}_{y'}
                            +\hat{\sigma}_3 \mathbf{e}_z.
\end{equation}
With the help of Eqs. (\ref{varpi'-s3}) and (\ref{SP'}), it becomes
\begin{equation*}
    \mathbf{s}'_0=s_{01} (\mathbf{e}_{x} \cos \phi -\mathbf{e}_{y} \sin \phi )
                 +s_{02} (\mathbf{e}_{x} \sin \phi +\mathbf{e}_{y} \cos \phi )
                 +s_{03}  \mathbf{e}_{z},
\end{equation*}
which is related to the polarization (\ref{s}) in the old gauge by
\begin{equation}\label{GT-PV}
    \mathbf{s}'_0=\exp[i (\hat{\mathbf \Sigma} \cdot \mathbf{e}_z) \phi ] \mathbf{s}_0.
\end{equation}
In a word, \emph{the polarization is a gauge-dependent quantity}. Eq. (\ref{GT-PV}) is the gauge transformation on the polarization corresponding to the gauge transformation (\ref{JV'}) on the PWF.
More importantly, the degree of freedom to specify the gauge has physically observable effects.

\textit{Observable effect of the gauge degree of freedom}
We have seen that only in a particular gauge can the polarization be represented by the Pauli matrices (\ref{PM}). This means, on the basis of the canonical quantum condition (\ref{CCR}), that only in a particular gauge can the polarization be canonically quantized.
But because the quantum condition (\ref{CCR}) is independent of the gauge, the polarization quantum number that follows from this quantum condition and the corresponding eigen PWF are physically meaningless unless they are associated with a particular gauge.
It is thus concluded from Eqs. (\ref{a}) and (\ref{s}) that one can change the polarization of the photon by changing the gauge of the PWF with the PWF itself remaining fixed.
Specifically, if one changes the gauge of a given PWF $\tilde{a}_0$ from the old to the new one, one will change the electric vector from (\ref{a}) into
\begin{equation}\label{aR}
    \mathbf{a}_0^R=\varpi'_0 \tilde{a}_0.
\end{equation}
Taking Eq. (\ref{varpi'}) into account, one has
\begin{equation*}
    \mathbf{a}_0^R=\exp[-i(\hat{\mathbf \Sigma} \cdot \mathbf{e}_z)\phi] \mathbf{a}_0.
\end{equation*}
Changing the gauge of a PWF amounts to rotating the corresponding electric vector about the momentum. In principle, $\mathbf{a}_0^R$ does not have the same polarization as $\mathbf{a}_0$ does.
Considering that the polarization is gauge dependent, it is proper to compare polarizations between $\mathbf{a}_0^R$ and $\mathbf{a}_0$ in a same gauge. For this purpose, one substitutes Eq. (\ref{varpi'-s3}) into Eq. (\ref{aR}) to give
$
\mathbf{a}_0^R=\varpi_0 \tilde{a}_0^R
$,
where
\begin{equation}\label{rot-Ta}
    \tilde{a}_0^R=\exp (-i \hat{\sigma}_3 \phi) \tilde{a}_0
\end{equation}
is the PWF of $\mathbf{a}_0^R$ in the old gauge. As a result, its polarization in the old gauge is
\begin{equation*}
    \mathbf{s}_0^R=\tilde{a}_0^{R\dag} \hat{\boldsymbol \sigma} \tilde{a}_0^R
                  =(s_{01} \cos 2\phi -s_{02} \sin 2\phi ) \mathbf{e}_x
                  +(s_{01} \sin 2\phi +s_{02} \cos 2\phi ) \mathbf{e}_y
                  + s_{03} \mathbf{e}_z,
\end{equation*}
which is related to the polarization (\ref{s}) of $\mathbf{a}_0$ in the same gauge by
\begin{equation}\label{rot-s}
    \mathbf{s}_0^R= \exp[-2i(\hat{\mathbf \Sigma} \cdot \mathbf{e}_z)\phi] \mathbf{s}_0.
\end{equation}
This is one observable effect of the gauge degree of freedom. It is observed that Eq. (\ref{rot-Ta}) is a rotation transformation on the PWF generated by $\hat{\sigma}_3$ in the old gauge. Eq. (\ref{rot-s}) shows that the polarization in the old gauge transforms as a vector under such a rotation.

\textit{Entanglement of the polarization with the momentum}
Now we are ready to find out a representation for the gauge degree of freedom by extending our results to a general quantum state and to explain its physical meaning. The Schr\"{o}dinger equation (\ref{SE-V}) has the following general solution,
\begin{equation*}
    \mathbf{f}(\mathbf{k}, t)=\mathbf{F}(\mathbf{k}) \exp(-i \omega t).
\end{equation*}
To pay attention to its polarization, it is convenient to introduce the unit VWF
$\mathbf{a} (\mathbf{k})=\mathbf{F} (\mathbf{k})/|\mathbf{F} (\mathbf{k})|$
that satisfies
\begin{equation}\label{norm1}
    \mathbf{a}^{\dag} \mathbf{a} =1.
\end{equation}
For the plane-wave state considered above, the unit VWF is
$\mathbf{a}=\mathbf{a}_0 \delta^3 (\mathbf{k}-k_0 \mathbf{e}_z)$.
With the unit VWF, the constraint (\ref{TC}) assumes
\begin{equation}\label{TC-a}
    \mathbf{a} \cdot \mathbf{w}=0.
\end{equation}
It means that the VWF $\mathbf a$ at each value of $\mathbf k$ can be expanded in terms of two orthogonal $\mathbf k$-dependent base vectors.
Let be $\mathbf u$ and $\mathbf v$ a pair of mutually perpendicular unit vectors that form with $\mathbf w$ a right-handed Cartesian reference system
$\mathbf{uvw}$, satisfying
\begin{equation}\label{triad}
    \mathbf{u} \times \mathbf{v} =\mathbf{w}, \quad
    \mathbf{v} \times \mathbf{w} =\mathbf{u}, \quad
    \mathbf{w} \times \mathbf{u} =\mathbf{v}.
\end{equation}
Choosing $\mathbf u$ and $\mathbf v$ as the base vectors, one can expand $\mathbf a$ as
$
\mathbf{a} =a_1 \mathbf{u} +a_2 \mathbf{v}
$.
The expansion coefficients at each value of $\mathbf k$ make up the corresponding Jones vector
$
    \tilde{a}=\bigg(\begin{array}{c}
                      a_1 \\
                      a_2
                    \end{array}
              \bigg)
$
in terms of which one can rewrite the expansion as
\begin{equation}\label{QUT-1}
    \mathbf{a} =\varpi \tilde{a},
\end{equation}
where
\begin{equation}\label{varpi}
    \varpi=(\begin{array}{cc}
              \mathbf{u} & \mathbf{v}
            \end{array}
           ).
\end{equation}
The $3 \times 2$ matrix $\varpi$ in Eq. (\ref{QUT-1}) performs a quasi unitary transformation in the following sense.
On one hand, $\varpi$ acts on a Jones vector to produce a VWF that satisfies the condition (\ref{TC-a}). It is easy to prove that
\begin{equation}\label{unitarity-2}
    \varpi^{\dag} \varpi =I_2.
\end{equation}
As a result, substituting Eq. (\ref{QUT-1}) into Eq. (\ref{norm1}) gives
\begin{equation}\label{norm2}
    \tilde{a}^{\dag} \tilde{a} =1.
\end{equation}
On the other hand, multiplying both sides of Eq. (\ref{QUT-1}) by $\varpi^{\dag}$ from the left and using Eq. (\ref{unitarity-2}), one has
\begin{equation}\label{QUT-2}
    \tilde{a} =\varpi^\dag \mathbf{a}.
\end{equation}
It says that $\varpi^{\dag}$ acts on a VWF to produce a Jones vector. Substituting it into Eq. (\ref{norm2}) and considering Eq. (\ref{norm1}), one is led to
\begin{equation}\label{unitarity-1}
    \varpi \varpi^{\dag} =I_3,
\end{equation}
where $I_3$ is the 3-by-3 unit matrix.
Eqs. (\ref{unitarity-2}) and (\ref{unitarity-1}) express the quasi unitarity \cite{Golub} of the matrix $\varpi$. $\varpi^{\dag}$ is the Moore-Penrose pseudo inverse of $\varpi$, and vice versa.

The same as the polarization (\ref{s}) of the photon in a plane-wave state, the polarization in the general state $\mathbf a$ is defined by
\begin{equation}\label{PV}
    \mathbf{s}=\tilde{a}^\dag \hat{\boldsymbol \varsigma} \tilde{a}
              =s_1 \mathbf{u} +s_2 \mathbf{v} +s_3 \mathbf{w},
\end{equation}
where
\begin{equation}\label{varsigma}
    \hat{\boldsymbol \varsigma}
   =\hat{\sigma}_1 \mathbf{u}+\hat{\sigma}_2 \mathbf{v}+\hat{\sigma}_3 \mathbf{w}
\end{equation}
and
$s_i=\tilde{a}^\dag \hat{\sigma}_i \tilde{a}$.
By this it is meant that the Jones vector (\ref{QUT-2}) plays the role of the PWF in such a way that the Pauli matrices (\ref{PM}) represent the Cartesian components of the polarization (\ref{PV}) in the local reference system $\mathbf{uvw}$.
Since the matrix $\varpi$ guarantees that the VWF (\ref{QUT-1}) satisfies the condition (\ref{TC-a}), the PWF (\ref{QUT-2}) is not constrained by such a condition. In other words, its two components are independent of each other.
But it is important to note that Eqs. (\ref{triad}) cannot completely determine the transverse axes, $\mathbf u$ and $\mathbf v$, of the local reference system $\mathbf{uvw}$ up to a rotation about the wavevector \cite{Mandel}.
That is to say, whether the PWF (\ref{QUT-2}) or the polarization operator (\ref{varsigma}) has not yet been completely determined.
To determine the polarization (\ref{PV}), one has to figure out a way to specify the transverse axes.
Fortunately, it has been shown in classical optics \cite{Stra, Green, Patt} that this can be done by introducing a constant unit vector, denoted by $\mathbf I$, as follows,
\begin{equation}\label{basis}
    \mathbf{u}=\mathbf{v} \times \frac{\mathbf k}{k},             \quad
    \mathbf{v}=\frac{\mathbf{I} \times\mathbf{k}}{|\mathbf{I} \times\mathbf{k}|}.
\end{equation}
Let us see how the polarization (\ref{PV}) is dependent on this vector.

Suppose that the transverse axes are specified by a new constant unit vector, $\mathbf{I}'$ say, as
\begin{equation*}
    \mathbf{u}'=\mathbf{v}' \times \frac{\mathbf k}{k},   \quad
    \mathbf{v}'=\frac{\mathbf{I}' \times\mathbf{k}}
                     {|\mathbf{I}' \times \mathbf{k}|}.
\end{equation*}
In this case, the VWF $\mathbf a$ is expressed similarly as
\begin{equation}\label{QUT-1'}
    \mathbf{a}=\varpi' \tilde{a}',
\end{equation}
where
$
\varpi'=(\begin{array}{cc}
            \mathbf{u}' & \mathbf{v}'
          \end{array}
        )
$
and
\begin{equation}\label{QUT-2'}
    \tilde{a}'=\varpi'^{\dag} \mathbf{a}
\end{equation}
is the new PWF.
As remarked earlier, the new transverse axes,
$\mathbf{u}'$ and $\mathbf{v}'$,
are related to the old ones, $\mathbf{u}$ and $\mathbf{v}$, by a rotation about $\mathbf k$.
Letting be $\Phi(\mathbf{k}; \mathbf{I}, \mathbf{I}')$ the $\mathbf k$-dependent rotation angle, such a rotation can be expressed as
\begin{equation}\label{pi-rotated}
    \varpi'=\exp [-i (\hat{\mathbf \Sigma} \cdot \mathbf{w}) \Phi] \varpi
\end{equation}
or as
\begin{equation}\label{rotation-pi}
    \varpi'=\varpi \exp [-i (\hat{\boldsymbol \varsigma} \cdot \mathbf{w}) \Phi],
\end{equation}
where $\hat{\boldsymbol \varsigma}$ is given by Eq. (\ref{varsigma}). Substituting Eq. (\ref{rotation-pi}) into Eq. (\ref{QUT-2'}) and considering Eq. (\ref{QUT-2}), one finds
\begin{equation}\label{GT-f}
    \tilde{a}'=\exp [i (\hat{\boldsymbol \varsigma} \cdot \mathbf{w}) \Phi] \tilde{a}.
\end{equation}
From these discussions it is concluded that the PWF (\ref{QUT-2}) is a gauge representation. The gauge is the local reference system $\mathbf{uvw}$. The degree of freedom to specify the gauge is the constant unit vector $\mathbf I$ that is introduced in Eqs. (\ref{basis}). Eq. (\ref{GT-f}) is the gauge transformation on the PWF.
In view of this, the polarization in the new gauge is given by
\begin{equation}\label{s'}
    \mathbf{s}'=\tilde{a}'^\dag \hat{\boldsymbol \varsigma}' \tilde{a}',
\end{equation}
where
\begin{equation*}
    \hat{\boldsymbol \varsigma}'
   =\hat{\sigma}_1 \mathbf{u}' +\hat{\sigma}_2 \mathbf{v}' +\hat{\sigma}_3 \mathbf{w}
\end{equation*}
is the corresponding polarization operator.
Substituting Eq. (\ref{GT-f}) into Eq. (\ref{s'}) and taking Eq. (\ref{rotation-pi}) into account, one finally gets
\begin{equation*}
    \mathbf{s}'=s_1 (\mathbf{u} \cos \Phi -\mathbf{v} \sin \Phi)
               +s_2 (\mathbf{u} \sin \Phi +\mathbf{v} \cos \Phi)
               +s_3 \mathbf{w},
\end{equation*}
which is related to the polarization (\ref{PV}) in the old gauge by
\begin{equation}\label{GT-s}
    \mathbf{s}'=\exp[ i (\hat{\mathbf \Sigma} \cdot \mathbf{w}) \Phi] \mathbf{s}.
\end{equation}
This shows that the polarization is gauge dependent. Eq. (\ref{GT-s}) is the gauge transformation on the polarization corresponding to the gauge transformation (\ref{GT-f}) on the PWF.

One of the observable effects of the gauge degree of freedom in the general case can be demonstrated in the same way as in the case of a plane wave. In particular, if the gauge of a given PWF $\tilde a$ is changed from the old to the new one, the VWF will be changed from (\ref{QUT-1}) into
$\mathbf{a}^R=\varpi' \tilde{a}$.
With the help of Eq. (\ref{pi-rotated}), it takes the form
\begin{equation}\label{rot-a}
    \mathbf{a}^R=\exp [-i (\hat{\mathbf \Sigma} \cdot \mathbf{w}) \Phi] \mathbf{a}.
\end{equation}
The polarization of $\mathbf{a}^R$ is of course different from that of $\mathbf a$. This is physically observable. Again \emph{changing the gauge of a PWF amounts to rotating the corresponding VWF about the momentum}. The so-called spin Hall effect \cite{Hosten} of the photon is such a physical process \cite{Li}.

The physical meaning of the gauge degree of freedom $\mathbf I$ can be understood as follows.
Even though it is usually compared to the spin of the electron in quantum mechanics \cite{Sakurai}, the polarization of the photon is sharply different from the spin of the electron. The spin of a free electron is supposed to be independent of its extrinsic variables such as the position and momentum. But Eq. (\ref{varsigma}) reveals that the polarization of the photon is entangled with its momentum in the sense that the Pauli matrices (\ref{PM}) represent the Cartesian components of the polarization in the local reference system $\mathbf{uvw}$ rather than in the laboratory reference system.
Now that it specifies the local reference system $\mathbf{uvw}$ or its transverse axes, the gauge degree of freedom determines how the polarization is entangled with the momentum.
Based on the conjugate relation between the position and momentum that is expressed by Eq. (\ref{E-vec}), it can be expected that the gauge degree of freedom also determines the entanglement of the polarization with the position \cite{Li}.
It is hoped that the results reported here will find important applications in the theory of quantum optics and in the techniques involving the polarization of the photon.

%\end{linenumbers}


\begin{thebibliography}{99}

\bibitem{Jauch} J. M. Jauch and F. Rohrlich, {\it The Theory of Photons and Electrons}, 2nd ed. (Springer-Verlag, New York, 1976).
\bibitem{Born} M. Born and E. Wolf, {\it Principles of Optics}, 7th ed. (Cambridge University Press, Cambridge, 1999).
\bibitem{Damask} J. N. Damask, {\it Polarization Optics in Telecommunications} (Springer, New York, 2005).
\bibitem{Kara} V. P. Karassiov, J. Phys. A {\bf 26}, 4345 (1993).
\bibitem{Luis} A. Luis and L. L. S\'{a}nchez-Soto, Prog. Opt. {\bf 41}, 421 (2000).
\bibitem{Marq} Ch. Marquardt, J. Heersink, R. Dong, M.V. Chekhova, A. B. Klimov, L. L. S\'{a}nchez-Soto, U. L. Andersen, and G. Leuchs, Phys. Rev. Lett. {\bf 99}, 220401 (2007).
\bibitem{Sode} J. S\"{o}derholm, G. Bj\"{o}rk, A. B. Klimov, L. L. S\'{a}nchez-Soto, and G. Leuchs, New J. Phys. {\bf 14}, 115014 (2012).
\bibitem{Hall} D. G. Hall, Opt. Lett. {\bf 21}, 9 (1996).
\bibitem{Stal} M. Stalder and M. Schadt, Opt. Lett. {\bf 21}, 1948 (1996).
\bibitem{Beck} A. M. Beckley, T. G. Brown, and M. A. Alonso, Opt. Express {\bf 18}, 10777 (2010).
\bibitem{Card} F. Cardano, E. Karimi, S. Slussarenko, L. Marrucci, C. de Lisio, and E. Santamato, Appl. Opt. {\bf 51}, C1 (2012).
\bibitem{Gadw} B. R. Gadway, E. J. Galvez, and F. De Zela, J. Phys. B {\bf 42}, 015503 (2009).
\bibitem{Kari} E. Karimi, J. Leach, S. Slussarenko, B. Piccirillo, L. Marrucci, L. Chen, W. She, S. Franke-Arnold, M. J. Padgett, and E. Santamato, Phys. Rev. A {\bf 82}, 022115 (2010).
\bibitem{Naga} E. Nagali, F. Sciarrino, F. De Martini, L. Marrucci, B. Piccirillo, E. Karimi, and E. Santamato, Phys. Rev. Lett. {\bf 103}, 013601 (2009).
\bibitem{Vall} A. Vall\'{e}s, V. D'Ambrosio, M. Hendrych, M. Mi\v{c}uda, L. Marrucci, F. Sciarrino, and J. P. Torres, Phys. Rev. A {\bf 90}, 052326 (2014).
\bibitem{D'Am} V. D'Ambrosio, G. Carvacho, F. Graffitti, C. Vitelli, B. Piccirillo, L. Marrucci, and F. Sciarrino, Phys. Rev. A {\bf 94}, 030304(R) (2016).
\bibitem{Pohl} D. Pohl, Appl. Phys. Lett. {\bf 20}, 266 (1972).
\bibitem{Jord-H} R. H. Jordan and D. G. Hall, Opt. Lett. {\bf 19}, 427 (1994).
\bibitem{Youn-B} K. S. Youngworth and T. G. Brown, Opt. Express {\bf 7}, 77 (2000).
\bibitem{Zhan} Q. Zhan, Adv. Opt. Photon {\bf 1}, 1 (2009).
\bibitem{Akhiezer} A. I. Akhiezer and V. B. Berestetskii, {\it Quantum Electrodynamics} (Interscience, New York, 1965).
\bibitem{Jauch-P} J. M. Jauch and C. Piron, Helv. Phys. Acta {\bf 40}, 559 (1967).
\bibitem{Amrein} W. O. Amrein, Helv. Phys. Acta {\bf 42}, 149 (1969).
\bibitem{Pauli} W. Pauli, {\it General Principles of Quantum Mechanics} (Springer-Verlag, New York, 1980).
\bibitem{Rose-S} D. Rosewarne and S. Sarkar, Quantum Opt. {\bf 4}, 405 (1992).
\bibitem{Biru} I. Bialynicki-Birula, in {\it Progress in Optics}, edited by E. Wolf (Elsevier, Amsterdam, 1996), Vol. 36, pp. 245-294.
\bibitem{Sakurai} J. J. Sakurai, {\it Modern Quantum Mechanics} (Benjamin/Cummings, California, 1985).
\bibitem{Cohen} C. Cohen-Tannoudji, J. Dupont-Roc, and G. Grynberg, {\it Photons and Atoms} (John Wiley \& Sons, New York, 1989).
\bibitem{Merz} E. Merzbacher, {\it Quantum Mechanics}, 3rd ed. (John Wiley \& Sons, New York, 1998).
\bibitem{Golub} G. H. Golub and C. F. Van Loan, {\it Matrix Computations}, 3rd ed. (Johns Hopkins, Baltimore, 1996).
\bibitem{Mandel} L. Mandel and E. Wolf, {\it Optical Coherence and Quantum Optics} (Cambridge University Press, New York, 1995).
\bibitem{Stra} J. A. Stratton, {\it Electromagnetic Theory} (McGraw-Hill, New York, 1941).
\bibitem{Green} H. S. Green and E. Wolf, Proc. Phys. Soc. A {\bf 66}, 1129 (1953).
\bibitem{Patt} D. N. Pattanayak and G. P. Agrawal, Phys. Rev. A {\bf 22}, 1159 (1980).
\bibitem{Hosten} O. Hosten and P. Kwiat, Science {\bf 319}, 787 (2008).
\bibitem{Li} C.-F. Li, Phys. Rev. A {\bf 79}, 053819 (2009).

\end{thebibliography}
\end{document}